\documentclass[a4paper,10pt,twoside]{cpc-hepnp}
\usepackage{amssymb}

\usepackage{multicol}
\usepackage{graphicx}
\usepackage{booktabs}
\usepackage{amssymb,bm,mathrsfs,bbm,amscd}
\usepackage[tbtags]{amsmath}
\usepackage{lastpage}
\usepackage{CJK}
\usepackage{color}

\begin{document}
\begin{CJK*}{GBK}{song}

\fancyhead[c]{\small Chinese Physics C~~~Vol. XX, No. X (2013)
XXXXXX} \fancyfoot[C]{\small 010201-\thepage}

\footnotetext[0]{Received 15 Dec 2013} \footnotetext[0]{*Supported
by National Science Foundation of
China(11205183,11005117,11225525,11390384 )}

\title{Temperature dependence of the light yield of the LAB-based and mesitylene-based liquid scintillators}

\author{%
      XIA Dong-Mei$^{1,}$$^{2}$\email{xiadm@ihep.ac.cn}%
\quad YU Bo-Xiang$^{1}$\email{yubx@ihep.ac.cn}%
\quad Li Xiao-Bo$^{1}$%
\quad SUN Xi-Lei$^{1}$%
\\
\quad DING Ya-Yun$^{1}$%
\quad ZHOU Li$^{1}$%
\quad CAO Jun$^{1}$%
\quad HU Wei$^{1,}$$^{2}$%
\quad YE Xing-Cheng$^{3}$%
\\
\quad CHEN Hai-Tao$^{4}$%
\quad DING Xue-Feng$^{3}$%
\quad DU Bing$^{1}$ \quad }

\maketitle

\address{%
$^1$ {\bf } (State Key Laboratory of Particle Detection and Electronics (Institute of High Energy Physics, CAS), Beijing, 100049, China)\\
$^2$ {\bf } (University of Chinese Academy of Sciences, Beijing, 100049, China)\\
$^3$ {\bf } (Wuhan University, Hubei, 430072, China)\\
$^4$ {\bf } (Nanjing University of Aeronautics and Astronautics, Jiangsu, 210016, China)\\
}

\begin{abstract}
We studied the temperature dependence of the light yield of the
linear alkyl benzene (LAB)-based and mesitylene-based liquid
scintillators. The light yield increases by 23\% for both liquid
scintillators when the temperature is lowered from $26\;^{\circ}$C to $-40\;^{\circ }$C,
correcting for the temperature response of the photomultiplier tube.
The measurements help to understand the energy response of the
liquid scintillator detectors.
Especially, the next generation reactor neutrino experiments for
neutrino mass hierarchy, such as the Jiangmen Underground Neutrino
Observatory (JUNO), require very high energy resolution. As no
apparent degradation on the liquid scintillator transparency was
observed, lowering the operation temperature of the detector to $\sim4\;^\circ$C will increase the photoelectron yield of the detector by 13\%, combining the light
yield increase of the liquid scintillator and the quantum efficiency
increase of the photomultiplier tubes.
\end{abstract}

\begin{keyword}
liquid scintillator, reactor neutrino, linear alkyl benzene, mesitylene
\end{keyword}

\begin{pacs}
 29.40.Mc
\end{pacs}

\footnotetext[0]{\hspace*{-3mm}\raisebox{0.3ex}{$\scriptstyle\copyright$}2013
Chinese Physical Society and the Institute of High Energy Physics of
the Chinese Academy of Sciences and the Institute
of Modern Physics of the Chinese Academy of Sciences and IOP Publishing Ltd}%

\begin{multicols}{2}

\section{Introduction}
Organic liquid scintillator (LS) is widely used to detect reactor neutrinos~\cite{kamland,dayabay,doublechooz,reno,borexino} due to its high light yield and high hydrogen fraction. Liquid scintillator is made of a solvent and a small amount of fluor, and often with an additional tiny amount of wavelength shifter. For example, the Daya Bay undoped liquid scintillator consists of linear alkyl benzene (LAB) as the solvent, 3 g/L 2,5-diphenyloxazole (PPO) as the fluor, and 15 mg/L p-bis-(o-methylstyryl)-benzene (bis-MSB) as the wavelength shifter, while the gadolinium-doped LS has the same recipe but mixed with a Gd complex with 0.1\% Gd in mass~\cite{dingyy,dybls}.

The energy response of the liquid scintillator detector need be well understood for precision measurements in a reactor neutrino experiment. The light yield of the liquid scintillator, to which the visible energy of an event in the detector is proportional, is however temperature dependent because of the thermal quenching effects. Excited solvent molecules by ionization may undergo non-radiation transition when colliding with other molecules. Normally the light yield will increase at lower temperature, when the viscosity of the solvent rises thus collisions reduce. If quencher presents in the solution, the situation will be complex since the collisions may also impede the energy transfer from the exited solvent molecule to the quencher.

The temperature dependence of the light yield has been studied for some liquid scintillators~\cite{buontempo} but is still scanty in literature. In this study, we will study the temperature effects of the Daya Bay LS, which is based on the relative new solvent LAB, over a range from $-40\;^{\circ }$C to $26\;^{\circ}$C. As a comparison, LS with the same solute fractions but another solvent, mesitylene, is also measured.

Such studies are of particular interests for the design of the next generation reactor neutrino experiments such as the Jiangmen Underground Neutrino Observatory (JUNO). To determine the neutrino mass hierarchy by precisely measuring the energy spectrum of the reactor neutrinos, JUNO detector requires a very high energy resolution of 3\%/$\sqrt{E({\rm MeV})}$~\cite{juno}. Previous experiments reached (5-6)\%/$\sqrt{E}$~\cite{kamland,borexino}. This unprecedent energy resolution requirement is a major challenge for JUNO. The increasing light yield of the LS at lower temperature provides an option to operate the detector at low temperature, e.g. at $\sim4\;^\circ$C, right above the ice point of the buffer water shielding the neutrino detector. For this purpose, we also studied the light transmittance of the LS at low temperature.

In this paper, the experimental setup, light yield measurement, correction for the temperature effects of the photomultiplier tube (PMT) are described in section 2. The transmittance is studied in section 3, followed by a conclusion and discussion.

\section{Temperature dependence of the light yield }

\subsection{Experimental setup}

The light yield of the LS is measured via Compton scattering of $\gamma$ rays from a radioactive source. To improve the precision of the measurement, we tag the scattered $\gamma$s at a fixed direction. The coincidence of the recoil electron in the LS and the scattered $\gamma$ selects the events of known deposited energy in the LS, thus reduces the uncertainty of the light yield measurement to sub-percent level from $\sim$5\% of the common method by fitting the Compton edge.

\begin{center}
\includegraphics[width=7cm]{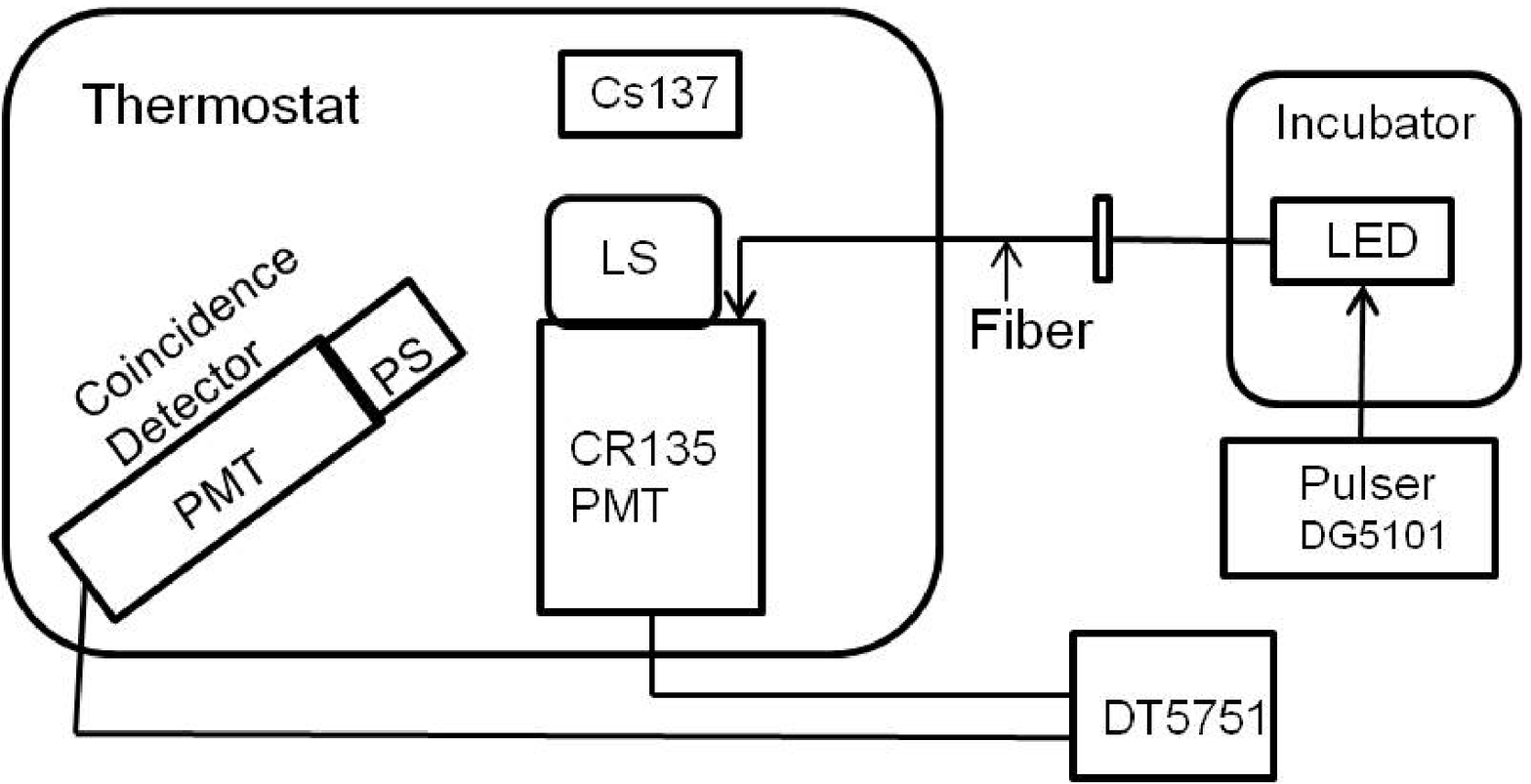}
\figcaption{\label{fig1}  Experimental setup for measuring the
temperature dependence of the LS light yield.}
\end{center}

Figure 1 shows the scheme of the experimental setup. The LS sample is contained in a cylindric quartz glass
vessel of 5 cm in diameter and 5 cm in height. The vessel is wrapped with Enhanced Specular Reflector (ESR) to increase the photon collection efficiency, and is coupled to a CR135 PMT. The scattered $\gamma$
is tagged by a coincidence detector, which is a plastic scintillator (PS) detector located about 10 cm from the LS at an angle of about $30^{\circ}$. The LS is irradiated by $\gamma$-rays from a $^{137}$Cs (15 $\mu$Ci) source. The source, the LS vessel, the CR135 PMT, and the coincidence detector are put in an enclosed thermostat with temperature adjustable from $-70\;^{\circ}$C to $155\;^{\circ}$C. The signals from the CR135 PMT and the coincidence detector are recorded by a CAEN DT5751 ADC unit with a self-trigger function. The relative light yield of the LS is determined by comparing the peak values measured by the ADC at different temperatures.

To separate the temperature effects of the PMT from that of the LS, a temperature-resistant optical fibre is coupled to the photocathode of the CR135 PMT directly, transmitting light from an LED driven by a pulse generator. The LED is located in an incubator box operating at 25 $^{\circ}$C, emitting light at 430 nm wavelength. The pulse generator is at room temperature, which is almost a constant during the measurement. The LED flashes at a frequency of 10 kHz during the LS light yield measurement and the data is triggered by the coincidence with the PS detector (noise). The probability of the LED signal overlapping with a Compton signal is found to be small enough.

The LS sample is bubbled with nitrogen from the bottom of the vessel before measurement to remove oxygen and water in the LS. Oxygen is a quencher of the LS. The presence of oxygen reduces the light yield of the LAB-based Daya Bay LS by up to 11\%~\cite{xiaohl}, shortens the time constant of the scintillation~\cite{lixb}, and may change the temperature effects of the LS. Normally the LAB-based LS contains tens ppm water. To avoid possible impacts on the light transmittance of the LS at low temperature, water is also removed by bubbling nitrogen. The LS sample is covered with nitrogen during the measurement. A temperature sensor is mounted on the LS vessel for monitoring. Data are taken only after the temperature has reached stable for more than 30 minutes.

\subsection{Relative light yield of the LS}

Figure 2 shows an example of the ADC distribution of the measurement. The light yield of the LS and the light intensity of the LED are measured by the ADC peaks, which are fitted with Gaussian functions. The LED intensity is stable to 1\% at constant temperature. The observed shift at different temperatures shows that the CR135 PMT response suffers from the temperature variation.

Figure 3 (a) shows the relative light yield of the LAB-based LS
before correcting for the PMT temperature effects. The measurements
have been done three times by different people and with slightly
different hardware. The data labelled "Third" corresponds to this
measurement and the other two were done before. The three measurements
are in good agreement. The temperature response of the CR135 PMT
monitored by the LED is shown in Figure 3 (b).

\begin{center}
\includegraphics[width=7cm]{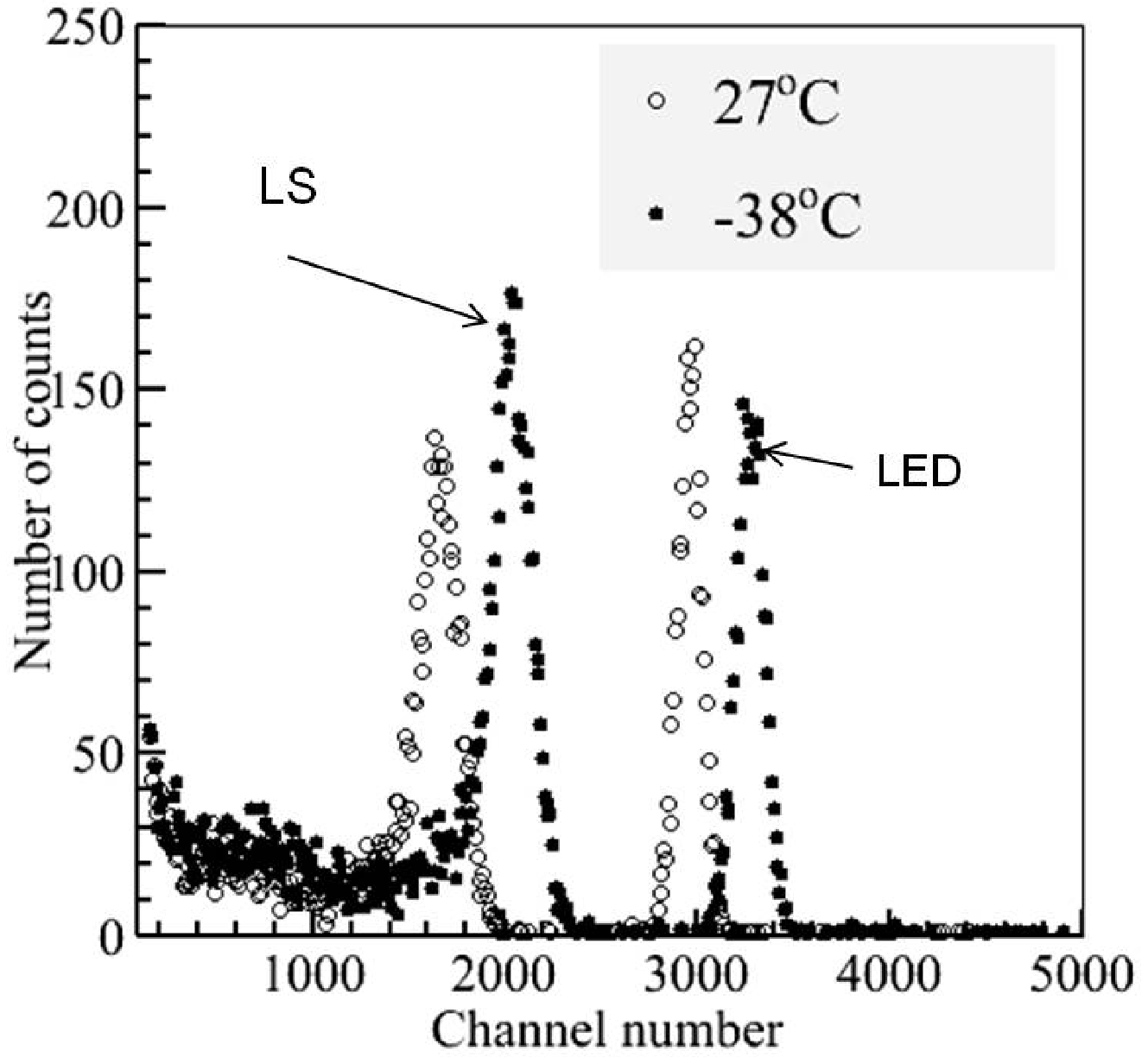}
\figcaption{\label{fig2}  ADC channel distribution of the LS events
and the LED monitoring signal.}
\end{center}

\begin{center}
\includegraphics[width=7cm]{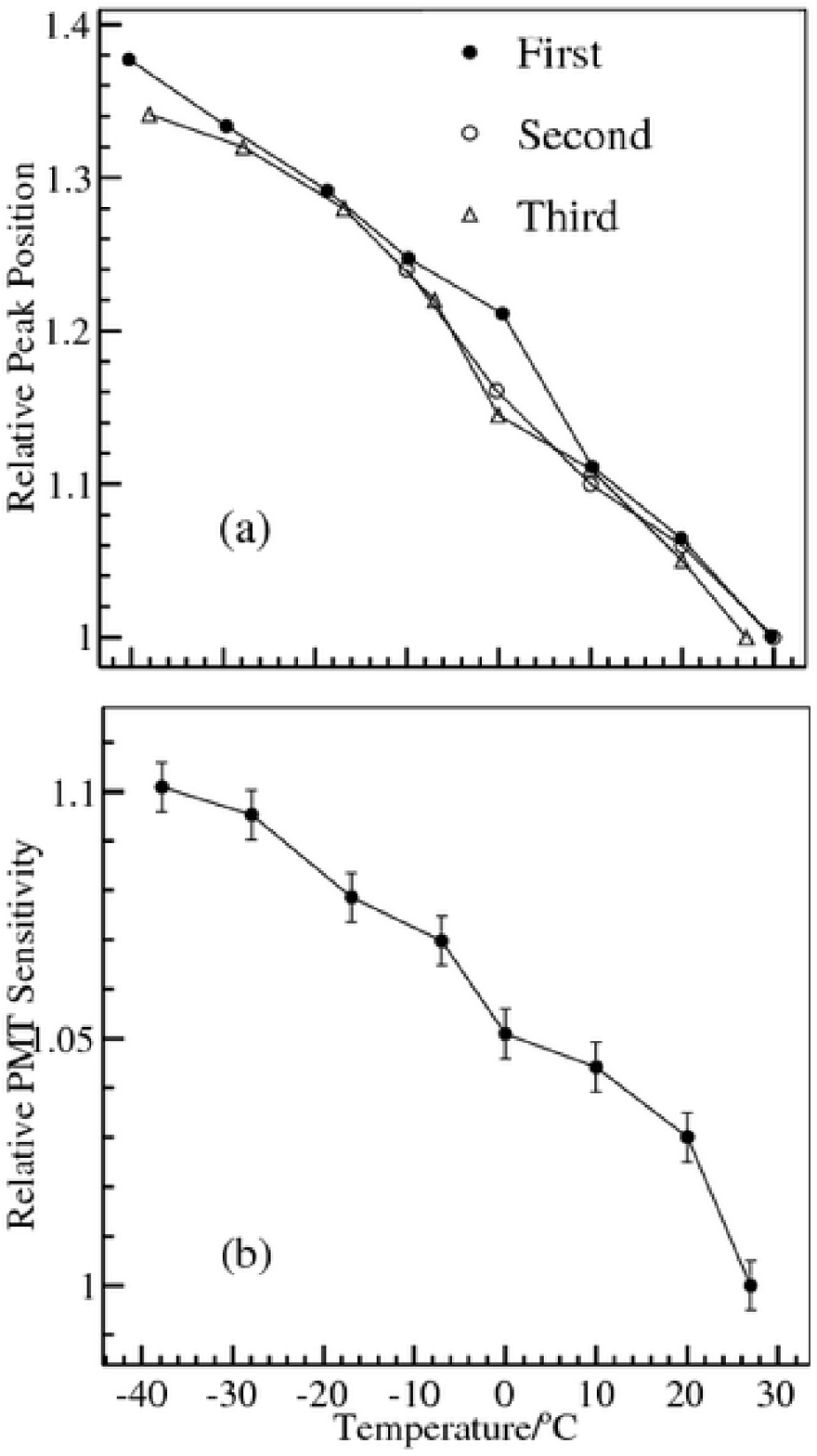}
\figcaption{\label{fig3}   (a) The relative light yield of the LAB-based LS before correcting for the PMT temperature effects, normalized at $26\;^\circ$C. (b) The temperature response of the CR135 PMT, normalized at $26\;^{\circ}$C.}
\end{center}

A similar measurement was done for the mesitylene-based LS. After correcting for the measured temperature
response of the PMT shown in Figure 3 (b), the temperature dependence of the LS light yield are shown in figure 4. The light yield increases by 23\% as the temperature decreases from $26\;^{\circ}$C to $-40\;^{\circ}$C for
both liquid scintillators.

\begin{center}
\includegraphics[width=7cm]{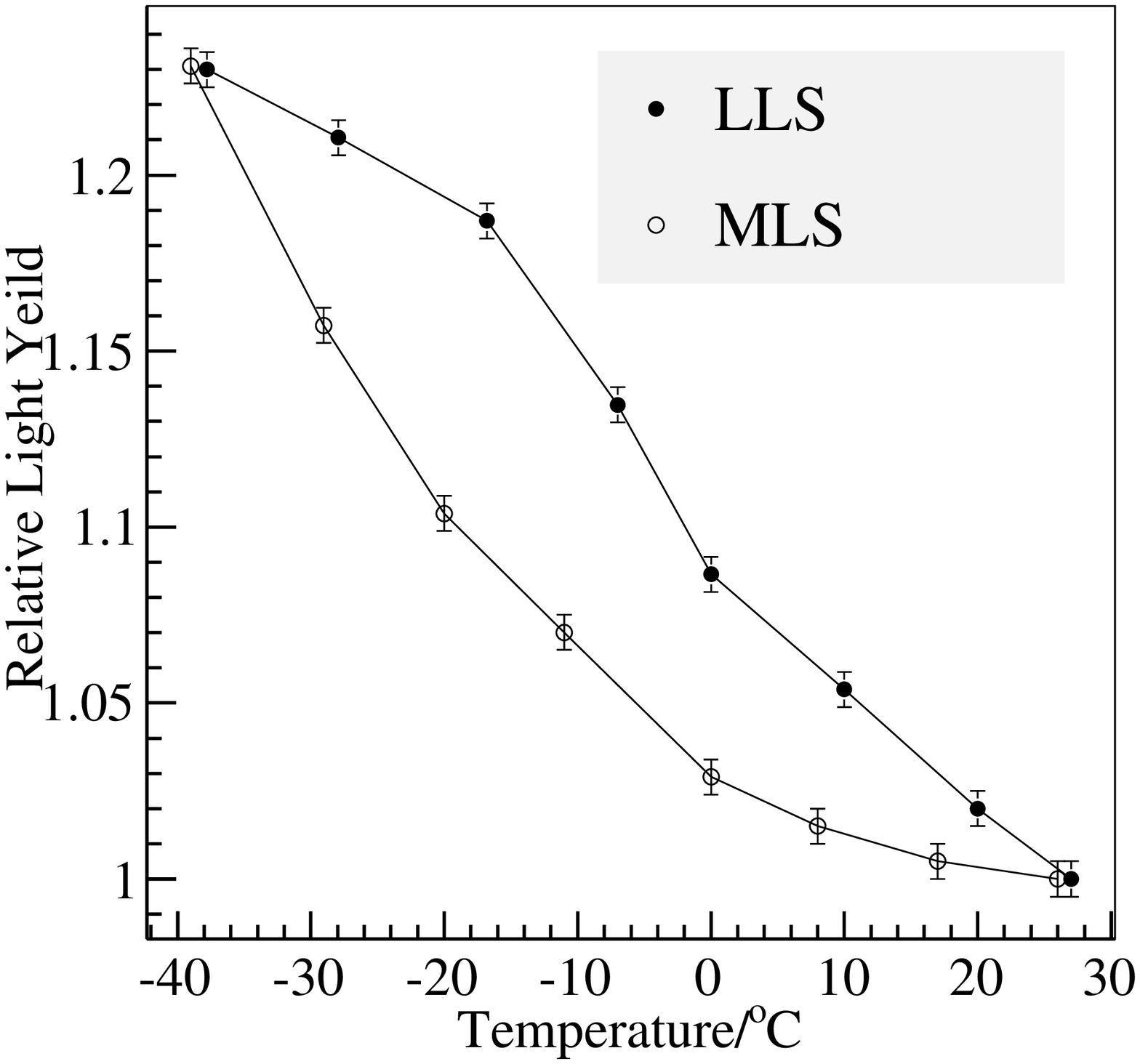}
\figcaption{\label{fig4}   The relative light yield of the LAB-based LS (LLS) and the mesitylene-based LS (MLS) after correcting for the PMT temperature effects, normalized at $26\;^{\circ}$C.}
\end{center}

The variation of the PMT response at difference temperature might be a
combination of the variation of the PTM gain and the quantum efficiency changes of the
PMT photocathode. The PMT gain decreases as the temperature increases because of
the negative temperature coefficient of the dynodes~\cite{mosz}.
Quantum efficiency of the PMT is typically 25\% at 430 nm at room
temperature for the CR135, which uses bialkali (SB-K-Cs) photocathode.
The quantum efficiency of the bialkali has almost a constant temperature coefficient of -0.2\%/$^{\circ}$C for photons of wavelength between 200 nm and 550 nm~\cite{hamamatsu}. It will increase by 13\% when the temperature is lowered from
$26\;^{\circ }$C to $-40\;^{\circ}$C.  Therefore, the quantum efficiency increase dominates the PMT temperature effects we have measured. In this measurement, The temperature stability of the temperature-resistant optical fibre is
estimated to be 0.5\% and the stability of the LED intensity is about 1\%.

\section{Temperature dependence of the light transmittance}

\subsection{Experimental setup}

For a large detector of $\sim$ 38 m diameter of JUNO, the light transmittance of the LS is equally important as the light yield. The temperature dependence of the light transmittance is
measured with 430 nm wavelength light, over a range from $-40\;^{\circ}$C to $26\;^{\circ}$C.

The scheme of the experimental setup is shown in Figure 5. The light source is a DH-2000 deuterium tungsten halogen lamp, followed by a monochromatic filter at 430 nm. The light transmits in a temperature-resistant fibre, passes the LS sample in a cuvette of dimension of
1 cm $\times$ 1 cm $\times$ 3 cm, and is received by an Ocean Optics QE65000 spectrophotometer. A temperature
probe is attached to the cuvette to monitor the LS temperature. The spectrum of the light is shown in the bottom panel of figure 5, measured by the spectrophotometer. The relative transmittance can be obtained by continuously measuring the light intense as temperature varies.

The surface of the cuvette may frost and bias the transmittance measurement when the temperature goes down. We observed such phenomenon that the transmittance started to drop dramatically at certain temperature (e.g. $-7\;^\circ$C for one of our measurements of LAB-based LS), although it is not visible on the surface of the cuvette by eye. After excluding the possible causes such as the crystallization of water content in the LAB, precipitation of scintillation fluor, and freezing of the solvent itself, we improved the experimental setup by sealing the LS cuvette in a transparent
airtight box to against the frost. The box is flushed with dry nitrogen before the
experiment to remove the vapor in the air, and maintain a small positive pressure with nitrogen. The nitrogen is released from the box through a bubbler to monitor the airtightness of the box.

\begin{center}
\includegraphics[width=7cm]{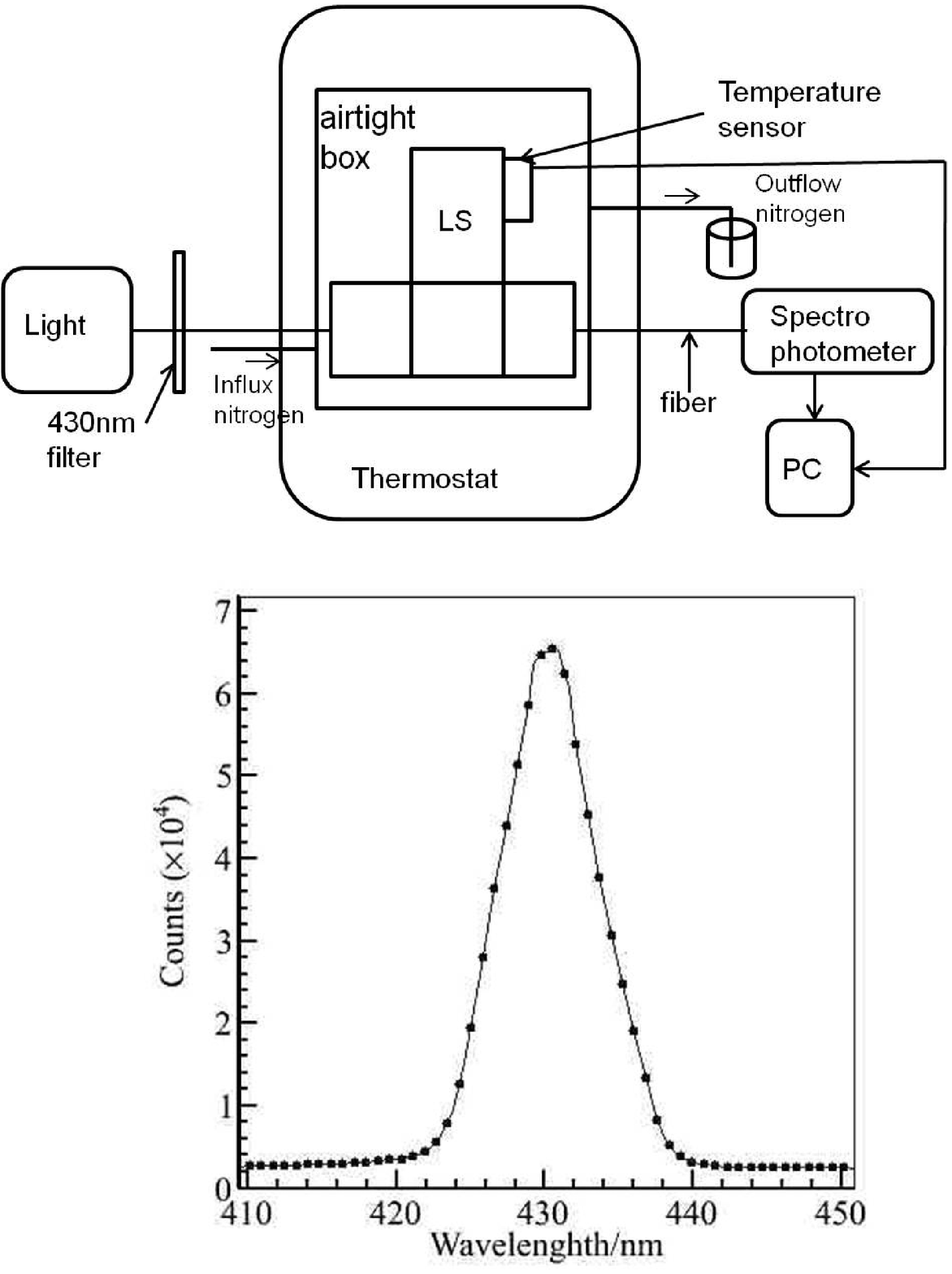}
\figcaption{\label{fig5}  Top: Schematic view of the experimental
setup for the light transmittance measurement. Bottom: The measured light spectrum after passing the monochromatic filter.}
\end{center}

\subsection{Relative transmittance of the LS}

Liquid scintillator exposed to air normally contains water of tens ppm.
Water crystals could be formed in the LS and degrade the transmittance at low temperature.
Water can be removed from the LS thoroughly by bubbling enough dry nitrogen. The nitrogen bubbling is also necessary to purge the oxygen in the LS. The water content in the LS can be measured to below 1 ppm with a 831KF Coulometric Moisture Analyzer. Table 1 shows the relationship between the
remaining water content and the volume of nitrogen flushed. Three litres of nitrogen is needed to purge all the water in the cuvette.

\begin{center}
\tabcaption{ \label{tab1}Relationship between the water content in the LS samples and the volume of nitrogen flushed.}
\footnotesize
\begin{tabular*}{80mm}{c@{\extracolsep{\fill}}ccc}
\toprule LS & 0 L   & 1 L  & 3 L \\
\hline
LLS\hphantom{00} & \hphantom{0} $\sim$ 27 ppm & \hphantom{0} $\sim$ 16 ppm & $\sim$ 0 ppm \\
MLS\hphantom{00} & \hphantom{0} $\sim$ 140 ppm & \hphantom{0} $\sim$ 70 ppm & $\sim$ 0 ppm \\
\bottomrule
\end{tabular*}
\end{center}

\par
Figure 6 presents the result of the relative transmittance for the LAB-based
and the mysitylene-based LS. We use the temperature
instability of the system without LS in the cuvette as the systematic error of the measurement.
For both the LAB-based and the mysitylene-based LS, transmittance stays stable when the
temperature decreases from  $26\;^{\circ}$C to $-40\;^{\circ}$C. The melting point of the LAB we used is below $-60\;^{\circ}$C and that of the mesitylene is $-44\;^{\circ}$C. Lowering the temperature to $-40\;^{\circ}$C will not cause a phase transition of the LS.

\begin{center}
\includegraphics[width=7cm]{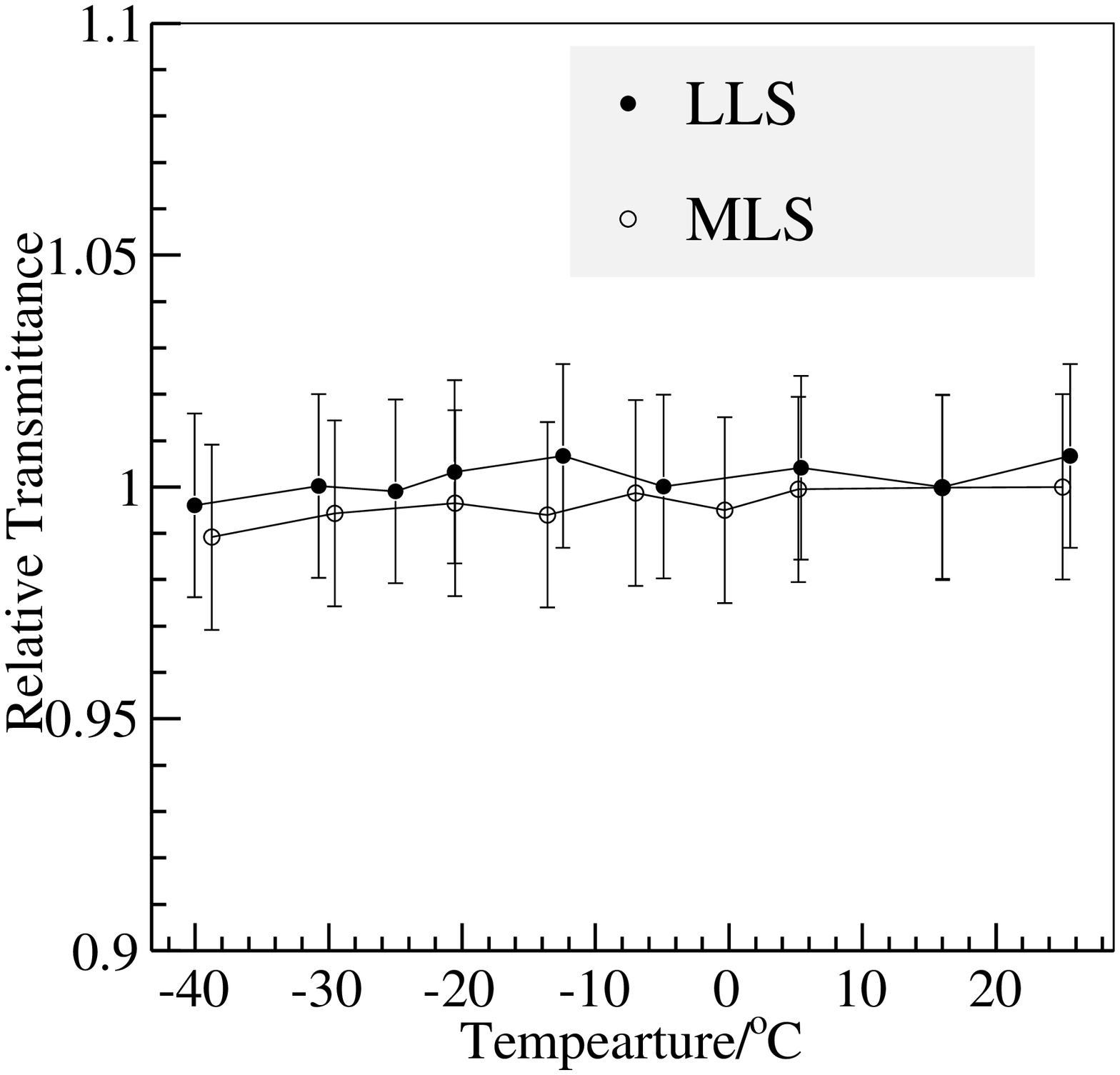}
\figcaption{\label{fig6}  Relative transmittance of the LAB-based LS (LLS) and the mysitylene-based LS (MLS) for light of 430 nm wavelength.}
\end{center}

\section{Conclusion and discussion}

The temperature dependence of the light yield and the transmittance of two liquid scintillators, LAB-based LS and
mysitylene-based LS, have been measured. For both liquid scintillators, when the temperature
is lowered from $26\;^{\circ}$C to $-40\;^{\circ}$C, the light
yield increases by 23\%, with the PMT effects corrected, and the transmittance remains stable. Frosting on the sample vessel at low temperature is observed to have significant impacts on the measurements. It is avoid by coupling the PMT to the sample vessel with silicone oil for the light yield measurement, or by putting the vessel in a nitrogen purged transparent box for the transmittance measurement.

The light yield increase at low temperature provides an option for the next generation reactor neutrino experiments for neutrino mass hierarchy such as JUNO, which requires very high energy resolution. When using water as outer buffer, the operation temperature of the detector could be lowered to $\sim 4\;^\circ$C. As no apparent degradation on
the liquid scintillator transparency was observed, lowering the
operation temperature to $\sim 4\;^\circ$C from $26\;^\circ$C will increase the
photoelectron yield of the detector by 13\%, in which 9\% is from the light yield of the LAB-based liquid scintillator and $\sim 4$\% is from the quantum efficiency of the PMT with bialkali photocathode.

Operating at even lower temperature, such $-40\;^\circ$C, will increase the photoelectron yield by 32\%. But it is more difficult to realize and requires an oil buffer instead of a water buffer.

\end{multicols}

\begin{multicols}{2}

\end{multicols}

\vspace{-1mm} \centerline{\rule{80mm}{0.1pt}} \vspace{2mm}

\begin{multicols}{2}

\end{multicols}

\clearpage

\end{CJK*}

\begin{thebibliography}{13}

\vspace{3mm}

\bibitem{kamland} Eguchi K {\it et al.} (KamLAND collaboration), Phys. Rev. Lett., 2003, {\bf 90}: 021802.
\bibitem{dayabay} An F P {\it et al.} (Daya Bay collaboration), Phys. Rev. Lett., 2012, {\bf 108}: 171803; Chin. Phys. C, 2013, {\bf 37}(1): 011001.
\bibitem{doublechooz} Aberle C {\it et al.} Nucl. Phys. B (Proc. Suppl.),2012,{\bf 229}: 448.
\bibitem{reno} Park J S {\it et al.} Nucl. Instr. and Meth. A, 2013,{\bf 707}: 45.
\bibitem{borexino} Alimonti G {\it et al.} (Borexino collaboration), Astropart. Phys., 2002, {\bf 16}(3): 205.
\bibitem{dingyy} Ding Y Y {\it et al.}, Nucl. Instr. and Meth. A, 2008, {\bf 584}: 238.
\bibitem{dybls} Ding Y Y {\it et al.}, submitted to Nucl. Instr. and Meth. A.
\bibitem{buontempo} Buontempo S {\it et al.}, Nucl. Instr. and Meth. A, 1999, {\bf 425}: 492.
\bibitem{juno} Li Y F, Cao J, Wang Y F, and Zhan L, Phys. Rev. D, 2013, {\bf 88}: 013008.
\bibitem{xiaohl} XIAO H L {\it et al.}, Chin. Phys. C, 2010, {\bf 34}(05): 571.
\bibitem{lixb} LI X B {\it et al.}, Chin. Phys. C, 2011, {\bf 35}(11): 1026.
\bibitem{mosz} Moszynski M {\it et al.}, Nucl. Instr. and Meth. A, 2006, {\bf 568}: 739.
\bibitem{hamamatsu} {\it Photomultiplier Tubes: Basics and Applications (Third Edition)}, Hamamatsu Photonics K.K., Japan (2006).

\end{thebibliography}
\end{document}